\documentclass[runningheads]{llncs}
\usepackage{graphicx}
\usepackage{cite}
\usepackage{amsmath,amssymb,amsfonts}
\usepackage{algorithmic}
\usepackage{textcomp}
\usepackage{xcolor}
\usepackage{todonotes}
\usepackage{censor}
\usepackage{subcaption}
\usepackage{wrapfig}
\usepackage{amsmath}  % <---
\usepackage{makecell} % <---
	
\usepackage{color, colortbl}
\usepackage{arydshln}

\usepackage{enumitem}

\usepackage{soul}
\usepackage{url}
\usepackage{flushend}
\usepackage[most]{tcolorbox}

\usepackage{hyperref}
\hypersetup{
    colorlinks=true,
    linkcolor=blue,
    filecolor=blue,      
    urlcolor=blue,
    citecolor=blue
    }

\newtcolorbox{myquote}[1][]{%
    colback=black!5,
    colframe=black!5,
    notitle,
    sharp corners,
    borderline west={2pt}{0pt}{gray!80!black},
    enhanced,
    breakable,
    }
    
\definecolor{Gray}{gray}{0.9}    
\definecolor{Gray2}{gray}{0.8}
    
\newcommand{\fig}[1]{Fig.~\ref{fig:#1}}   
\newcommand{\tabl}[1]{Table~\ref{tab:#1}}

\newcommand{\footnoteref}[1]{%
    $^{\ref{#1}}$%
}

\begin{document}

%
% \title{End-to-End Test Coverage in Microservices}
% \title{On End-To-End Test Coverage for Microservices Endpoints}

\title{End-to-End Test Coverage Metrics in Microservice Systems: An Automated Approach}
\titlerunning{E2E Test Coverage Metrics in Microservice Systems}
%
%\titlerunning{Test Coverage Degradation in Microservice Systems}
\author{Amr S. Abdelfattah\inst{1}\orcidID{0000-0001-7702-0059} \and
Tomas Cerny\inst{2}\orcidID{0000-0002-5882-5502} \and
Jorge Yero Salazar\inst{1}\orcidID{0000-0002-5033-4805} \and
Austin Lehman\inst{1} \and
Joshua Hunter\inst{1} \and
Ashley Bickham\inst{1} \and
Davide Taibi\inst{3}\orcidID{0000-0002-3210-3990}}
\authorrunning{Abdelfattah et al.}
\institute{Computer Science, Baylor University, One Bear Place 97141 Waco, TX, USA \\
\email{amr\_elsayed1@baylor.edu} \and
Systems and Industrial Engineering, University of Arizona, Arizona, USA \\
\email{tcerny@arizona.edu} \and
University of Oulu, Oulu, Finland \\
\email{davide.taibi@oulu.fi}} 
% \\
% \email{lncs@springer.com}}
\maketitle              % typeset the header of the contribution
\begin{abstract}
Microservice architecture gains momentum by fueling systems with cloud-native benefits, scalability, and decentralized evolution. However, new challenges emerge for end-to-end (E2E) testing. Testers who see the decentralized system through the user interface might assume their tests are comprehensive, covering all middleware endpoints scattered across microservices. However, they do not have instruments to verify such assumptions.
This paper introduces test coverage metrics for evaluating the extent of E2E test suite coverage for microservice endpoints. Next, it presents an automated approach to compute these metrics to provide feedback on the completeness of E2E test suites. Furthermore, a visual perspective is provided to highlight test coverage across the system's microservices to guide on gaps in test suites.
We implement a proof-of-concept tool and perform a case study on a well-established system benchmark showing it can generate conclusive feedback on test suite coverage over system endpoints.

%eliminating the need for testers to have access to internal system details.
% This paper proposes test coverage metrics to verify the scope of E2E testing coverage with regard to microservice endpoints. Additionally, it provides an automated approach to calculate the metrics over systems without exposing testers to system internal details.
%to give feedback on the completeness of their test suites. Furthermore, they can use a visual approach that highlights test coverage across the system fragmented into microservices to guide testers on gaps in test suites.

 %, while unaware of the system's internal details.
 
%The approach considers the dynamic analysis of the test execution to generate traces from which we identify tested endpoints and pair them with particular tests. 

%In the context of decentrally evolving microservices, this new instrument brings critical infrastructure to mitigate test degradation~over~time.

%unaver of system internal details
\keywords{microservices \and end-to-end testing \and API tests \and test quality}
\end{abstract}
\setcounter{footnote}{0} 

\section{Introduction}
% \vspace{-0.5em}
Microservice architecture enables practitioners to build scalable software systems broken down into a collection of loosely coupled interacting services. Each service is responsible for a specific business capability and can be developed and deployed independently of other services. This allows for faster development cycles, easier maintenance, and better scalability.

% Testers focus on E2E system validation interacting with the system through its user interface, which hides the underlying logical system structure. Microservice architecture contains more details than the traditional monolithic systems architecture, such as multiple services and inter-dependencies, which add challenges in the E2E testing of microservice systems since it hides more details that could impact the testing completenss and efficiency.

However, the end-to-end testing of microservice systems can be challenging due to the system's distributed nature hidden from testers. During E2E system validation, testers primarily interact with the system through its user interface, thereby concealing the underlying logical system structure. However, microservice architecture entails more intricate details compared to traditional monolithic systems, including multiple services, inter-dependencies, and continuous evolution. Testers may lack knowledge about the specific services being involved and executed within the system. Consequently, they may encounter difficulties in testing all possible scenarios. This complexity introduces challenges in E2E testing of microservice systems, as it obscures crucial details that can influence testing completeness and efficiency.

% This situation is sketched in~\fig{intro}. Testers might not necessarily know the specific services being involved and executed. As a result, they may not be able to test all possible scenarios and may miss critical paths. This problem is compounded when the system comprises multiple microservices that interact with each other and continuously evolve.  

% \begin{figure}[htbp]

% \begin{wrapfigure}{r}{0.48\textwidth} 
% \vspace{-2.6em}
% \centering
%     \includegraphics[width=15em]{img/intro.pdf}
%     \vspace{-1em}
%     \caption{The E2E tester view of the systems}%\todo[inline]{Amr: update the image}} 
%     \label{fig:intro}
%     \vspace{-3em}
% \end{wrapfigure}
% \end{figure}

The extent to which a particular system's microservices are involved in individual E2E tests or E2E test suites should be recognized to give testers better insights into system coverage and test-to-microservice dependencies (i.e., test evolution). E2E tests interact with the system through the user interface which mediates the the interaction to microservice endpoint level. Thus, associating tests with impacted microservice endpoints they interact with would provide testers with insights into how comprehensive their test suites are when contrasted to all system endpoints.

This paper aims to establish metrics for calculating the coverage of endpoints in E2E test suites their individual tests, and microservices. Furthermore, it aims to propose a practical method and measurement approach through a case study. This work considers microservice endpoints as the points of overlap between the logical system structure and the E2E tests. It proposes an automated approach mapping individual tests to system microservices and their endpoints to guide testers in test design completeness. With the detailed knowledge of test-to-endpoint associations, testers can better understand their test suite coverage and identify unobvious gaps.

This paper makes the following contributions in the context of microservices:
\setlist{leftmargin=7.5mm}
\setlist[itemize]{ topsep=2pt}
\begin{itemize}
\item[\textbullet] Proposal of three metrics (Microservice endpoint coverage, Test case endpoint coverage, and Complete Test suite endpoint coverage) to assess the coverage of endpoints in E2E testing.
\item[\textbullet] Metric extraction process and proof-of-concept tool imlementation.
\item[\textbullet] A practical system case study deriving and validating the coverage metrics.
\end{itemize}

This paper elaborates on related work in Section 2 and describes the metrics and process in Section 3. A case study is detailed in Section 4 followed by a discussion in Section 5 and conclusions in Section 6.

\section{Related Work}
% \vspace{-0.5em}
% Testing and analyzing microservices through the use of endpoints is an important challenge in the software field. Since microservices are often viewed as a black box, the underlying complexity of the system is often left unknown. Finding approaches to analyze microservice systems and find the underlying services is a vital area of study that continues to be improved.

%%% Testing Importance

% Moreover, Waseem et al. \cite{waseem2021design} conducted a survey of microservice practitioners to understand how microservice systems are designed, monitored, and tested in the industry. In terms of understanding the decomposition of microservice systems, they found that domain-driven design and business capability were the main strategies used. 

%%SAR & Testing approaches

Various studies have identified the lack of assessment techniques for microservice systems. A systematic literature review by Ghani et al.~\cite{Ghani_2019} concluded that most articles focused on testing approaches for microservices lacked sufficient assessment and experimentation. Jiang et al.~\cite{jiang2022efficient} emphasized the need for improved test management in microservice systems to enhance their overall quality.

Waseem et al.~\cite{waseem2021design} conducted a survey and revealed that unit and E2E testing are the most commonly used strategies in the industry. However, the complexity of microservice systems presents challenges for their monitoring and testing, and there is currently no dedicated solution to address these issues. Similarly, Giamattei et al.~\cite{giamattei2022automated} identified the monitoring of internal APIs as a challenge in black box testing microservice systems, advocating for further research~in~this~area.

To address these gaps, it is crucial to develop an assistant tool that improves system testing and provides appropriate test coverage assessment methods. Corradini et al.~\cite{testing-metrics} conducted an empirical comparison of automated black-box test case generation approaches specifically for REST APIs. They proposed a test coverage framework that relies on the API interface description provided by the OpenAPI specification. Within their framework, they introduced a set of coverage metrics, consisting of eight metrics (five request-related and three response-related), which assess the coverage of a test suite by calculating the ratio of tested elements to the total number of elements defined in the API. However, these metrics do not align well with the unique characteristics of microservice systems. They do not take into account the specific features of microservices, such as inter-service calls and components like API gateway testing.

Giamattei et al.~\cite{giamattei2022automated} introduced MACROHIVE, a grey-box testing approach for microservices that automatically generates and executes test suites while analyzing the interactions among inter-service calls. Instead of using the commonly used tools such as SkyWalking or Jaeger, MACROHIVE builds its own infrastructure, which incurs additional overhead by requiring the deployment of a proxy for each microservice to monitor. It also involves implementing communication protocols for sending information packets during request-response collection. MACROHIVE employs combinatorial tests and measures the status code class and dependencies coverage of internal microservices. However, compared to our proposed approach, MACROHIVE lacks static analysis of service dependencies, relying solely on runtime data. In contrast, our approach extracts information statically from the source code, providing accurate measurements along with three levels of system coverage.

Ma et al.~\cite{ma2018using} utilized static analysis techniques and proposed the Graph-based Microservice Analysis and Testing (GMAT) approach. GMAT generates Service Dependency Graphs (SDG) to analyze the dependencies between microservices in the system. This approach enhances the understanding of interactions among different parts of the microservice system, supporting testing and development processes. GMAT leverages Swagger documentation to extract the SDG, and it traces service invocation chains from centralized system logs to identify successful and failed invocations. The GMAT approach calculates the coverage of service tests by determining the percentage of passed calls among all the calls, and it visually highlights failing tests by marking the corresponding dependency as yellow on the SDG. However, GMAT is tailored to test microservices using the Pact tool and its APIs. In contrast, our approach introduces three coverage metrics that focus on different levels of microservice system parts, emphasizing endpoints as fundamental elements of microservice interaction. While our approach doesn't consider the status code of each test, combining GMAT with our proposed approach could offer further insights for evaluating microservice testing and assessment criteria.

In summary, this paper tackles the gap in assessment techniques for microservice testing. It aims to introduce test coverage metrics and develop an analytical tool that can assess microservice systems and measure their~test~coverage.

\section{The E2E Test Coverage Metrics}
\vspace{-.5em}

% \todo[inline]{Amr: Could you create one more picture to detail the static analysis process as you have done for the Dynamic analysis. Then you can have consistent explanation in this section regarding four parts: static analysis, dynamic analysis, calculate the coverage by merging the sub-process results, then, the visualization approach to depict the coverage.}

% \todo[inline]{Amr: Please create this section regarding four parts: 1- static analysis, 2- dynamic analysis, 3- calculate the coverage by merging the sub-process results, 4- the visualization approach to depict the coverage. You need to describe the pictures related to each part.}

% This section details the methodology for our approach and its stages, as well as how we designed and implemented a prototype tool to reflect this approach.
 % and the design and implementation of a prototype tool that embodies this approach. 

 % Specifically, the approach aims at calculating the microservice endpoint coverage in tests.
 
% This section provides our proposed metrics and a comprehensive overview of our automated approach and its stages for applying the metrics. It measures how effectively the E2E testing suites cover the endpoints in microservices-based systems.

This section presents our proposed metrics and provides a comprehensive overview of our automated approach, outlining its stages for extracting the data required for calculating the metrics over systems. The objective is to assess E2E testing suites in achieving coverage of endpoints within microservices-based systems.

% In this section, we present a comprehensive overview of our approach, including its stages. 

\vspace{-.7em}
\subsection{The Proposed Metrics Calculations}

E2E testing involves test suites, where each test suite contains test cases that represent a series of steps or actions defining a specific test scenario. We introduce three metrics to assess the coverage of endpoints in microservice systems: microservice endpoint coverage, test case endpoint coverage, and complete test suite coverage. These metrics are described in detail below:
% \vspace{-0.5em}

 \setlist{leftmargin=4.5mm}
\begin{itemize}
    \item \textbf{Microservice endpoint coverage:} determines the tested endpoints within each microservice. It is obtained by dividing the number of tested endpoints from all tests by the total number of endpoints in that microservice. This metric offers insights into the comprehensiveness of coverage for individual microservices. The formula for microservice endpoint coverage~is:

    \vspace{-0.5em}
\begin{myquote}
\noindent
{\footnotesize
    % \[ C_{\text{ms}(i)} = \frac{E_{\text{ms}(i)}^{\text{tested}}}{E_{\text{ms}(i)}^{\text{total}}}  \quad \text{;} \quad\]
     \[ C_{\text{ms}(i)} = \frac{|E_{\text{ms}(i)}^{\text{tested}}|}{|E_{\text{ms}(i)}|}  \quad \text{;} \quad\]
    \[\begin{aligned}
        C_{\text{ms}(i)} & \text{- the coverage per microservice } i, \\
        E_{\text{ms}(i)}^{\text{tested}} & \text{ - the set of tested endpoints in microservice } i, \\
        E_{\text{ms}(i)} & \text{ - the set of all endpoints in microservice } i.
    \end{aligned}\]
 }   
 \end{myquote}
    % \[ coverage\_per\_ms_i = \frac{\#tested\_endpoints\_in\_ms_i}{\#total\_endpoints\_in\_ms_i} \]

    \item \textbf{Test case endpoint coverage:} gives a percentage of endpoints covered by each test case. It is calculated by dividing the number of endpoints covered by each test by the total number of endpoints in the system. This provides insights into the effectiveness of individual tests in covering the system's endpoints. The formula for test case endpoint coverage is:
    % \[ C_{\text{test}(i)} = \frac{E_{\text{test}(i)}^{\text{tested}}}{E_{\text{total}}^{\text{system}}}  \quad \text{;} \quad \]

    \vspace{-0.5em}
\begin{myquote}
\noindent
 {\footnotesize
    \[ C_{\text{test}(i)} = \frac{|E_{\text{test}(i)}^{\text{tested}}|}{ |\bigcup_{j}^{m\_total}{E_{\text{ms}(j)}}|}  \quad \text{;} \quad \]
    \[\begin{aligned}
    C_{\text{test}(i)} & \text{ - the coverage per test } i, \\
    E_{\text{test}(i)}^{\text{tested}} & \text{ - the set of tested endpoints from test } i, \\
    { m\_total } & \text{ - the total number of microservices in the system}, \\
    { \bigcup_{j}^{m\_total}{E_{\text{ms}(j)}}} & \text{ - the set of all endpoints in the system}.
    % E_{\text{total}}^{\text{system}} & \text{ represents the total number of endpoints in the system}.
\end{aligned}\]
}
\end{myquote}

    % \[ coverage\_per\_test_i = \frac{\#tested\_endpoints\_from\_test_i}{\#total\_endpoints\_in\_system} \]

    \item \textbf{Complete Test suite endpoint coverage:} determines the test suite overall coverage of the system by dividing the total number of unique endpoints covered by all tests by the total number of endpoints in the system. It provides insights into the completeness of test suites in covering all endpoints within the system. The formula for complete test suite endpoint coverage is:

    \vspace{-0.5em}

\begin{myquote}
\noindent
{\footnotesize
    % \[ C_{\text{suite}} = \frac{E_{\text{tested}}^{\text{all tests}}}{E_{\text{total}}^{\text{system}}} \quad \text{;} \quad \]
    \[ C_{\text{suite}} = \frac{|\bigcup_{i}^{t\_total}{E_{\text{test}(i)}^{tested}}|} {|\bigcup_{j}^{m\_total}{E_{\text{ms}(j)}}|} \quad \text{;} \quad \]
    
    \[\begin{aligned}
    C_{\text{suite}} & \text{ - the complete test suite coverage}, \\
    { m\_total } & \text{ - the total number of microservices in the system}, \\
    { t\_total } & \text{ - the total number of tests in the test suite}, \\
    \bigcup_{i}^{t\_total}{E_{\text{test}(i)}^{tested}} & \text{ - the set of all tested endpoints from all tests}, \\
    \bigcup_{j}^{m\_total}{E_{\text{ms}(j)}} & \text{ - the set of all endpoints in the system}.
\end{aligned}\]
}
\end{myquote}

    % \[ complete\_test\_suite\_coverage = \frac{\#tested\_endpoints\_from\_all\_tests}{\#total\_endpoints\_in\_system} \]
\end{itemize}
\begin{figure}[b!]
\vspace{-2em}
\centering
\includegraphics[width=0.7\columnwidth]{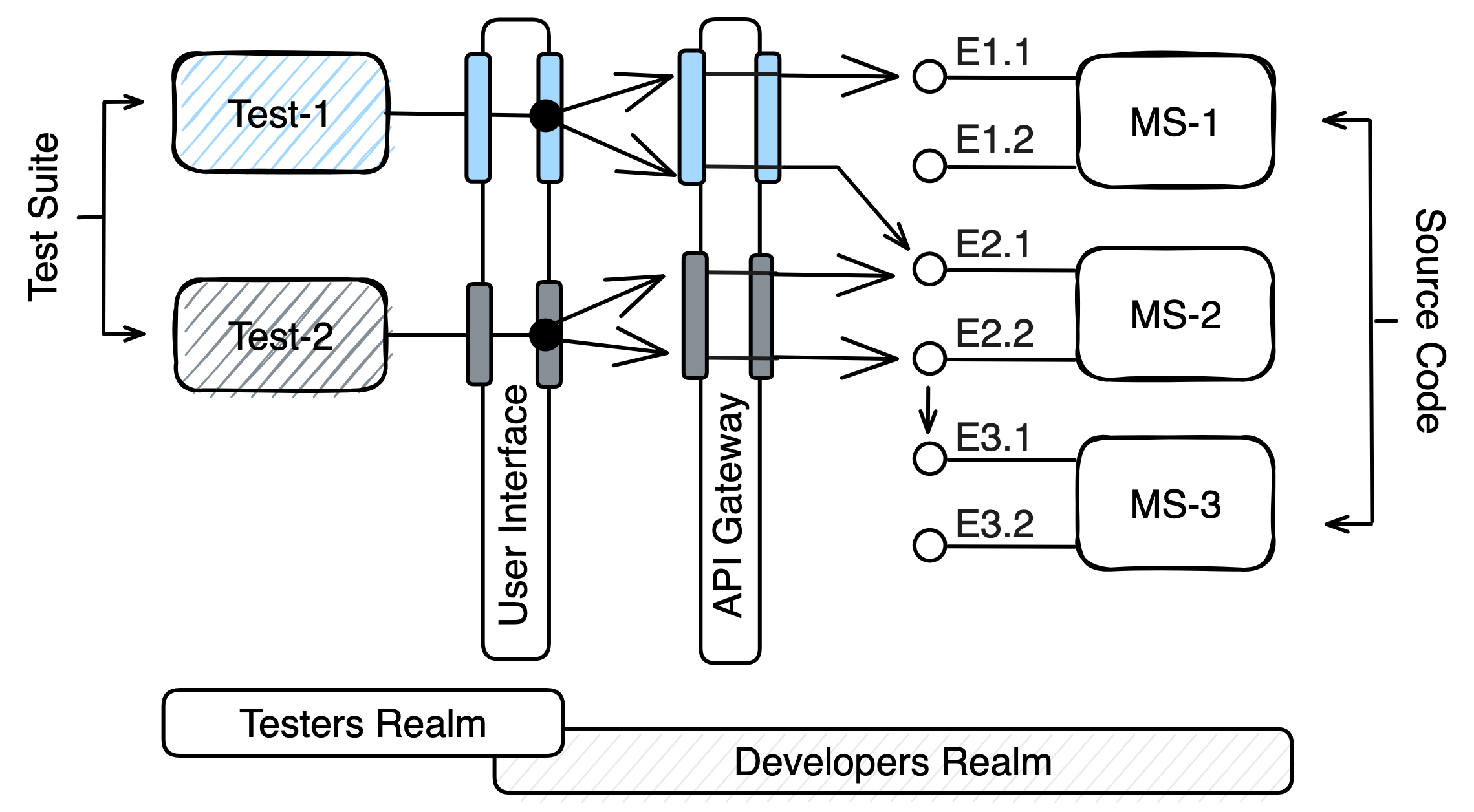}
\vspace{-1em}
\caption{Calculation Clarification Example}
\label{fig:example}
% \vspace{-2em}
\end{figure}
% \end{wrapfigure}

To provide further clarification, consider a system consisting of three microservices (MS-1, MS-2, MS-3), each with two endpoints, with a test suite composed of two tests (Test-1, Test-2), as depicted in~\fig{example}. In the example, the tests interact with endpoints through the user interface, which triggers the initiation of endpoint requests passed through the API gateway component. The example demonstrates that Test-1 calls two endpoints, one from MS-1 (E1.1) and one from MS-2 (E2.1). On the other hand, Test-2 calls two endpoints from MS-2 (E2.1, E2.2), E2.2 has an inter-service call to endpoint E3.1~in~MS-3.

% It demonstrates how the coverage equations are applied to calculate the microservice endpoint coverage, test case endpoint coverage, and complete test suite coverage. 

 % Test-1 interacts with the endpoints of MS-1 and MS-2. 

Applying our metrics, we can calculate the microservice endpoint coverage ($C_{\text{ms}(i)}$) for each microservice. For MS-1 and MS-3, only one out of their two endpoints is tested throughout all tests, resulting in a coverage of 50\% ($C_{\text{ms}(1)}=C_{\text{ms}(3)}=\frac{1}{2}$) for each. However, for MS-2, both of its endpoints are tested at least once, leading to a coverage of 100\% ($C_{\text{ms}(2)}=\frac{2}{2}$).

Next, we calculate the test case endpoint coverage ($C_{\text{test}(i)}$) per each test.Test-1 covers two out of the six endpoints in the system, resulting in a coverage of approximately 33.3\% ($C_{\text{test}(1)} = \frac{2}{6}$). Test-2 covers three distinct endpoints, resulting in a coverage of 50\% ($C_{\text{test}(2)} = \frac{3}{6}$). It is important to highlight that Test-2 contains an inter-service call to endpoint E3.1, which is considered in our approach.

Finally, we can calculate the complete test suite endpoint coverage ($C_{\text{suite}}$) of the system. Out of the six endpoints in the system, four distinct endpoints are tested from the two tests. This results in $\approx 66.6\%$ coverage ($C_{\text{suite}} = \frac{4}{6}$).

% \todo[inline]{Tomas: I consider the option to move this example to section 3.2 after the metrics. I }

\vspace{-1em}
\subsection{The Metrics Extraction Process}\label{sec:process}

To automatically collect the data for calculating the test coverage metrics, we propose to employ a combination of static and dynamic analysis methods. 

\begin{figure}[b!]
\centering
\vspace{-1.8em}
    \includegraphics[width=0.8\textwidth]{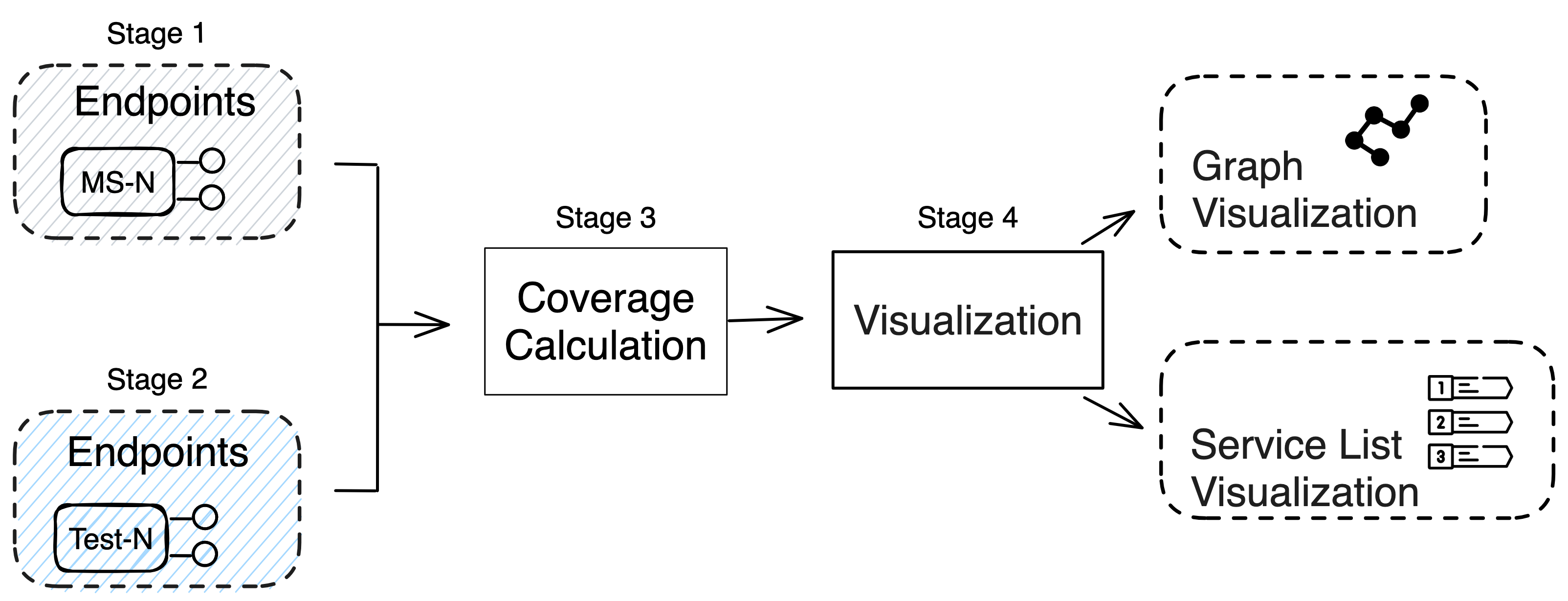}
       \vspace{-1em}
    \caption{The proposed approach overview}%\todo[inline]{Amr: update the image}} 
    % \todo[inline]{Amr: - Surround the top part with a rectangle called 'Static Analysis', and the left side with a rectangle called 'Dynamic Analysis' (DONE).  - Use those two parts to detail the methodology, explain how the static and dynamic analysis have done. - Replace 'User Interface' with 'visualization' and use different shape for it.  - Replace the first top 'Microservice' with `Microservice Source Code`.  - Make different shape for 'Coverage' to represent that is an output. - It's not clear what 'Rest Calls' means? I think it could be `System Logs`!}}
    \label{fig:approach-overview}
    \vspace{-.8em}
\end{figure}

The static analysis phase focuses on examining the source code to extract information about the implemented endpoints in the system. The dynamic analysis phase involves inspecting system logs and traces to identify the endpoints called by the automation tests. By combining the data obtained from both analyses, the approach applies the proposed metrics to generate the E2E endpoint coverage, and then it provides two visualization approaches to depict the coverage over the system representation. This process involves the following four~stages as illustrated in~\fig{approach-overview}:
\begin{itemize}[leftmargin=2cm]
    \item[\textbf{Stage 1.}] Endpoint Extraction From Source Code (Static Analysis).
    \item[\textbf{Stage 2.}] Endpoint Extraction From Log Traces (Dynamic Analysis).
    \item[\textbf{Stage 3.}] Coverage Calculation.
    \item[\textbf{Stage 4.}] Coverage Visualization. 
\end{itemize}

% The static analysis phase focuses on examining the source code to extract information about the implemented endpoints in the system. On the other hand, the dynamic analysis phase involves inspecting system logs and traces to identify the endpoints impacted by the automation tests. By combining the data obtained from both analyses, the approach generates the test coverage. Moreover, this approach offers two user-friendly visualization approaches. The first approach presents a list of microservices, where each microservice is accompanied by a list of its covered and uncovered endpoints. The second one employs a graph visualization, which provides a visual representation of the coverage percentage of each microservice along with its dependencies. This enables users to gain valuable insights into the level of test coverage achieved by the employed automated test suites within the microservices context.

% By analyzing various aspects of the system, including its codebase and runtime behavior, the approach aims to provide insights into the extent of test coverage achieved by test automation frameworks in the context of microservices.

We will delve into the details of each stage to demonstrate the approach.

% The approach aims at calculating the test coverage of testing automation frameworks (i.e., Selenium) in a microservice-based systems. It involves deta genegrated from static and dynamic analysis to the system.

%Our goal is to test on a real microservice system in order to visualize the coverage across different microservices. We also want to include the ability to view said services independently from each other as this is a feature not currently available with any other test coverage application.
%\todo[inline]{Amr: This paragraph is not a methodology related. The method doesn't know about 'real microservice system' Make sure that you briefly describe the purpose of the method and its components that you will detail in the rest of the section.}

% \begin{wrapfigure}{r}{0.6\textwidth}

\vspace{-1.5em}
\subsubsection{Stage 1: Expoint Extraction From Source Code (Static Analysis):}

% This phase focuses on analyzing the source code of microservices-based systems to extract the endpoints included inside its implementations.

% Static analysis is an analysis method performed on computer programs without executing them. It examines program code without running it. Analyzing the syntax and structure of the code it extracts information about the system~\cite{chess04}.

\begin{figure}
\centering
   % \vspace{-2.8em}
    \includegraphics[width=0.8\textwidth]{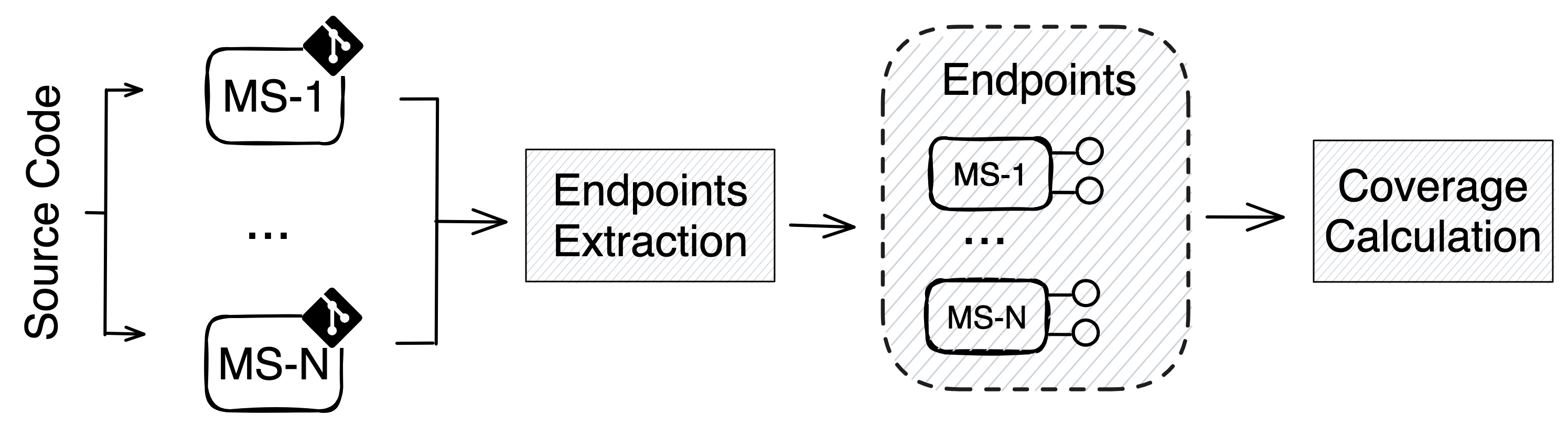}
       % \vspace{-2em}
       \vspace{-1em}
    \caption{Stage 1: Static analysis flow}
    
    % \todo[inline]{Amr: How does the static analysis generate System Logs? I think it needs correction.}}
    \label{fig:static-flow}
    \vspace{-1.8em}
    % \vspace{-2.5em}
\end{figure}

% Static analysis is analyzing the syntax and structure of the code without executing them to extract information about the system.
% and within the whole system ($E_{\text{total}}^{\text{system}}$)
Our approach applies a static analysis approach to the system's source code to extract the employed endpoints in each microservice ($E_{\text{ms}(i)}$). Static analysis refers to the process of analyzing the syntax and structure of code without executing it in order to extract information about the system. As depicted in~\fig{static-flow}, initially, microservices can be divided and detected from the system codebase. Each microservice's codebase is then processed by the \textit{endpoint extraction process}, which produces the endpoints corresponding to each microservice.

% \begin{wrapfigure}{r}{0.6\textwidth}
% \end{wrapfigure}

The identification of API endpoints typically relies on specific frameworks or libraries. For example, in the Java Spring framework, annotations such as \texttt{@RestController} and \texttt{@RequestMapping} are commonly used. This ensures consistency in metadata identification. Code analysis extracts metadata attributes about each endpoint, including the path, HTTP method, parameters, and return type. However, identification of endpoints can be performed across platforms as demonstrated by Schiewe et al. \cite{9737496} or accomplished by frameworks like Swagger\footnote{Swagger \url{https://swagger.io}}

As a result, a list of endpoints is generated and organized according to the respective microservice they belong to. This comprehensive list of endpoints becomes one of the inputs for our \textit{coverage calculation process}, where it combines the output of the dynamic analysis flow.

% First, the actual system code must be statically analyzed in order to generate a list of all of the endpoints grouped by microservices. This approach has minimal overhead and accurate endpoint identification. The importance of getting all of the endpoints is so that we can later properly identify which endpoints went untested by the provided test suite.

% \todo[inline]{SOMEONE CHECK}
% Figure 2 shows an expansion to the black box abstraction of the static analysis represented in Figure 1. When presented with the codebase, the system can be fragmented into various microservices. Each microservice codebase can be passed into an endpoint enumeration service that extracts all of the projects' microservices into the respective endpoints associated with each microservice. 

% As a result, a list of these endpoints will then be generated that is collected and sorted by the individual microservice, it is associated with. This is the total list of endpoints that is input to our coverage controller that will be compared against the dynamic test analysis.

% \subsection{Dynamic Flow Analysis of Test Execution Trace}

\vspace{-1em}
\subsubsection{Stage 2: Endpoint Extraction From Log Traces (Dynamic Analysis):}

\begin{figure}[h!]
\vspace{-2em}
    \includegraphics[width=\textwidth]{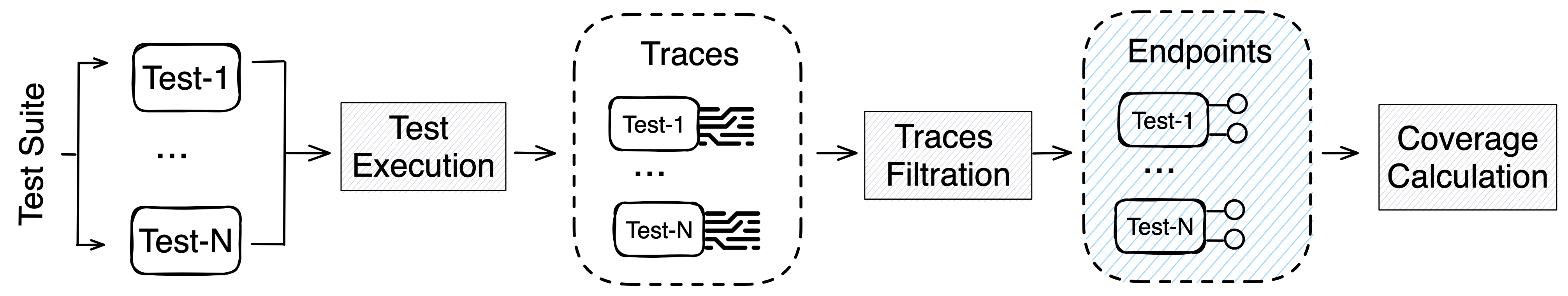}
       \vspace{-1.8em}
    \caption{Stage 2: Dynamic analysis flow}%\todo[inline]{Amr: update the image}}
    %\todo[inline]{Amr: It's not clear what it lead to! I suggest getting the all picture combined back without mentioning `Selenium`, just say `Dynamic Test Runner` that should work for any tool not only selenium. But, this picture in its own is fragmented, and the first picture is incomplete.}
    % }
    \label{fig:dynamic-flow}
    \vspace{-2em}
\end{figure}

% Dynamic analysis involves running the analyzed system to observe its runtime behavior and transactions. It provides insights into data flow, memory usage, performance, and potential issues. Dynamic analysis validates a system under real-world conditions. Microservices typically use centralized tracing to aid system debugging and monitoring. Collected trace logs can be used for this purpose.
% $E_{\text{tested}}^{\text{all tests}}$) from the test suite
We utilize dynamic analysis to identify the endpoints called during the execution of each test case in test suites ($E_{\text{test}(i)}^{\text{tested}}$. It also identifies the microservices containing these tested endpoints ($E_{\text{ms}(i)}^{\text{tested}}$). The analyzed system is executed to observe its runtime behavior and transactions. This analysis involves running multiple E2E tests and capturing the traces that occur, as illustrated in~\fig{dynamic-flow}.

The dynamic analysis flow sketched in~\fig{dynamic-flow} has two main responsibilities. Firstly, it takes the tests and executes them sequentially. During the execution of the E2E tests, traces are generated, capturing the interactions with the system. These traces are sent to a configured centralized logging system (i.e., SkyWalking, Jaeger), which stores them in its own storage, or an externally configured data storage solution (i.e., Elasticsearch), enabling analysis and further processing. Secondly, the process calculates the delta of the produced traces to identify the traces relevant to each executed test. This can be achieved in various ways, such as recording a timestamp from the start of a test's execution to its completion, retrieving the traces after each test execution and calculating the difference based on the latest track record, or sending a dynamically generated trace before and after the execution of each test to mark the start and end. In our approach, we have employed the first strategy, as it avoids unnecessary processing and complexity at this stage.

The extracted test trace sequences corresponding to each test undergo a \textit{traces filtration} process that filters and identifies the traces related to endpoints. This may involve queries to the trace storage to return specific trace indexes in the data. For instance, the SkyWalking tool marks the traces involving endpoint calls and makes them accessible under an index (in particular, \texttt{sw\_endpoint\_relation\_server\_side} index). Additionally, centralized logging systems encode the data records using \texttt{Base64\footnote{\label{foot-base64}Base64: \url{https://developer.mozilla.org/en-US/docs/Glossary/Base64}}} when sending them to external storage like Elasticsearch. Therefore, this step may include an additional decoding process if needed to detect the endpoints. These endpoint-related trace records contain information about the source and destination endpoints involved in the call relationship.

As a result, a list of endpoints is generated and organized according to the respective test suite they belong to. This list of endpoints becomes the second input for the \textit{coverage calculation process}, where it is combined with the output of the static analysis stage.

\vspace{-1em}
\subsubsection{Stage 3: Coverage Calculation:}

This stage combines the extracted equations from the previous two stages to calculate the three metrics of coverage ($C_{\text{ms}(i)},\ C_{\text{test}(i)},\ C_{\text{suite}}$). %both static and dynamic analyses. 

A challenge arises when matching the extracted system endpoints from the source code with those extracted from the traces. Since traces contain invoked endpoints with arguments' values, while those identified by static analysis hold parameter types and names. A similar challenge has been accounted for when profiling systems using log analysis and matching log lines with logging statements in the source code \cite{zhao2014lprof}. The source code contains a log message template with parameters, and execution logs contain a message with values from the execution context, which is not a direct match (i.e., source code \texttt{log.info('calling \{a\} from \{b\}')} vs. a contextual log statement \texttt{'calling for from bar'} where both \texttt{a} and \texttt{b} are interpreted). Zhao et al. have identified all code log statements to extract templates that could be matched using regular expressions to identify and match the parameter types whose values are present in the log output. 

In our approach, we employ signature matching to solve the challenge. It involves comparing the endpoint method signature with the data and parameters exchanged during REST calls communication to detect and verify the authenticity and matches of the requests. Thus, to determine which system endpoints were called by the test we consider the comparison of extracted attributes of the endpoints (such as path, request type, and parameter list) from the source code with the REST calls extracted from the test traces. This matching process helps to establish the coverage levels and determine which endpoints were effectively exercised by the tests.

% In this phase, the approach calculates the three mentioned levels of coverage and applies their equations based on the enumerated endpoints generated from both static and dynamic analyses. Constructing which of the source code endpoints got called from the tests, this approach used signature matching. It matches the extracted endpoints-related attribites: path, request type, and parameter list, from the source code endpoints and the endpoints extracted from the tests' logs. We followed a similar approach to the established logs comparison in~\cite{zhao2014lprof}, especificly for using regular expresssions to identfy and match the parameters types whose values appear in the log output.

\vspace{-1em}
\subsubsection{Stage 4: Coverage Visualization:}

The approach offers two ways to visualize these coverage metrics. The first displays a list of microservices, with each microservice showing its endpoints. Covered endpoints are marked in green, while missed endpoints are marked in red, as demonstrated in~\fig{vis-list}. The second representation utilizes the service dependency graph, where microservices are represented as nodes, and the dependencies between them are shown as edges. The nodes in the graph are color-coded based on the coverage percentage, allowing users to visually observe the coverage on the holistic system view depicting service dependencies, as exampled in~\fig{vis-graph}. These techniques help in visualizing the two metrics of $C_{\text{ms}(i)}$ and $C_{\text{test}(i)}$. Thus, these coverage calculations and visualizations provide valuable insights into the extent of test coverage achieved by automation frameworks in the context of microservices, enabling users to assess the effectiveness of their testing efforts and identify areas that require improvement.

\vspace{-1.2em}
\section{Case Study}
\vspace{-0.5em}

To demonstrate the completeness of our approach, we implemented a prototype and conducted a case study on an open-source system benchmark and an E2E test suite designed for the same system. We calculated our metrics on the testbench and compared the results with a manually calculated ground truth.
%
% Furthermore, we analyzed a Gatling-based load testing suite that aimed to achieve full endpoint coverage of the system. We evaluated the coverage achieved by our prototype in comparison to the coverage generated by the Gatling suite. This analysis provided insights into our prototype for achieving an effective test coverage of the system.

% To demonstrate our approach, we executed the implemented prototype over an open-source system benchmark and selenium test suite for the same system. We calculated the test coverage then validated the results of the experiment against a manually calculated ground truth. Additionally, we analyzed a gatling-based load testing suite that intended to produce full coverage of the system and evaluate our prototype coverage accordingly as well.

% When run the system benchmark and perform tests through our proof of concept tool to analyze the test coverage. We then manually verify the results of the experiment and indicate details specific to the system and the testing suite. 

% To demonstrate our approach, we considered an open-source system benchmark When run the system benchmark and perform tests through our proof of concept tool to analyze the test coverage. We then manually verify the results of the experiment and indicate details specific to the system and the testing suite. 

% \todo[inline]{Amr: I do expect this sections to contain subsections as: Proof of Concept Implementation, Benchmark Setup, Case Study Execution, and Results}

\vspace{-1em}
\subsection{Proof of Concept Implementation}

%\todo[inline]{Amr: Please move the implementation to the Case Study section DONE. And also, you need to give details about the }

This section describes the implementation of a prototype\footnote{\label{foot-prototype}Prototype: \url{https://github.com/cloudhubs/test-coverage-backend}} to showcase the four phases of the proposed approach. We focused on statically analyzing Java-based project source codes that use the Java Spring Cloud framework, an open-source framework that is widely used for building cloud-native applications. It provides developers with a comprehensive set of tools and libraries to build scalable and resilient applications in the Java ecosystem.

For the endpoint extraction from source code (Stage 1), we utilized the open-source JavaParser\footnote{\label{foot-javaparser}JavaParser: \url{https://github.com/javaparser/javaparser}} library. It allowed us to parse Java source code files, generate an Abstract Syntax Tree (AST) representation, and traverse it to detect spring annotations such as \texttt{@GetMapping} and \texttt{@PostMapping}. We extracted the relevant attributes once the endpoints were detected.

For the endpoint extraction from log traces (Stage 2), we utilized Apache Maven, a build automation tool for Java projects, to execute our JUnit test suites. JUnit,  a widely adopted unit testing framework, offers seamless integration with various automation test frameworks, including Selenium. On the other hand, we focused on extracting logs and traces from Elasticsearch, which is widely adopted as a central component in the ELK\footnote{\label{foot-elk}ELK: \url{https://aws.amazon.com/what-is/elk-stack}} (Elasticsearch, Logstash, Kibana) stack. We used the Elasticsearch Java High-Level REST Client\footnote{\label{foot-elastic-client}Elasticsearch Java Client: \url{https://www.elastic.co/guide/en/elasticsearch/client/java-rest/current/java-rest-high.html}}, which offers a convenient way to interact with Elasticsearch. It provided a QueryBuilder class to construct queries for searching and filtering data, such as creating a query to retrieve the logs that are between specific start and end timestamps.

Then, the prototype performs the coverage calculation (Stage 3). It integrates the results of the static and dynamic processes, and applies the proposed metrics. For the coverage visualization (Stage 4), we provided the two visualization approaches discussed earlier. We implemented a web application\footnote{\label{foot-prototype-ui}Coverage Visualizer: \url{https://github.com/cloudhubs/test-coverage-frontend}} that presents the information in an expandable list view for easy navigation. To integrate with the service dependency graph visualization, we utilized the Prophet library\footnote{\label{foot-prophet}Prophet: \url{https://github.com/cloudhubs/graal-prophet-utils}}, an open-source project that generates the graph from source code. Additionally, we utilized the visualizer library\footnote{\label{foot-visualizer}3D Visualizer: \url{https://github.com/cloudhubs/graal_mvp}}, which offers a tailored 3D microservices visualization for service dependency graphs.

\vspace{-1.2em}
\subsection{Benchmark and Test Suites}
\vspace{-0.5em}
To ensure unbiased testing of our application, we utilized an open-source testbench consisting of the TrainTicket system and associated test suites.

TrainTicket~\cite{trainticketgit} is a microservice-based train ticket booking system that is built using the Java Spring framework. It uses the standard annotations for defining the endpoints and uses the \textit{RestTemplate} Java client to initiate requests to endpoints. This benchmark consists of 41 Java-based microservices and makes use of Apache SkyWalking\footnote{\label{foot-skywalking}SkyWalking: \url{https://skywalking.apache.org/docs}} as its application performance monitoring system.

In order to run the TrainTicket system and execute tests on it, certain configuration fixes were necessary. To address this, a fork\footnote{\label{foot-trainticket}TrainTicket: \url{https://github.com/cloudhubs/train-ticket/tree/v1.0.1}} of the TrainTicket repository was created, specifically from the 1.0.0 release. This fork incorporated the necessary fixes and a deployment script. TrainTicket integrates with Elasticsearch, allowing our prototype to utilize SkyWalking for forwarding system logs to Elasticsearch for additional processing and analysis.

% TrainTicket~\cite{trainticketgit} is a microservice-based train ticket booking system implemented using the Java Spring framework. It comprises 41 Java-based microservices. It uses apache skywalking as an application performance monitoring (APM) system. To run the TrainTicket system and execute the tests on it, we made necessary configuration fixes and created a fork\footnote{\label{foot-trainticket}TrainTicket: https://github.com/cloudhubs/train-ticket/tree/1.0.1-release} from the 1.0.0 release, incorporating the required fixes, deployment script and Elastic Search integration to configure skywalking to forward the logs there for further processing through our prototype.

For the test suites, we utilized an open-source test benchmark\footnote{\label{foot-tests}Test benchmark: \url{https://github.com/cloudhubs/microservice-tests}} published in~\cite{smith2023benchmarks}. This benchmark aims to test the same version of the TrainTicket system. It contains 11 E2E test cases using the Selenium framework.

\vspace{-1em}
\subsection{Ground Truth}
\vspace{-0.5em}

To validate the completeness of our approach, we performed a manual analysis to construct the ground truth for the test benches. The complete results of the ground truth are published in an open accessed dataset\footnote{\label{foot-dataset}Dataset: \url{https://zenodo.org/record/8055457}}. This involved manual extraction of the data related to the first two stages in our proposed process in~section~\ref{sec:process}, as follows: endpoint extraction from source code and endpoint extraction from log traces.

For Stage 1, we validated the endpoints extracted during the static analysis by manually inspecting the source code of the microservices' controller classes. This allowed us to identify and extract information such as the endpoint's path, request type, parameter list, and return type. This process extracted 262 defined endpoints in the TrainTicket testbench codebase.

For Stage 2, we validated the endpoints extracted during the dynamic analysis by examining the Selenium test suites. Since the Selenium tests do not explicitly reference endpoints but rather perform UI-based actions, we manually analyzed the logs generated by the tests, which were stored in Elasticsearch. These logs contained encoded information about the source and destination endpoints, which we decoded and filtered to extract the list of endpoints called during the tests. It produced 171 unique endpoints from the logs.

\vspace{-1em}
\subsection{Case Study Results}
\vspace{-0.2em}
% We have started the execution by running the deployment script and get the TrainTicket system running on a local instance. After that, we executed our prototype over the test sutes to begin running the tests and generate the endpoints affected and calculate the test coverage as detailed in the methodology.

% The outcomes of the experiment execution is produced and published to this dataset\footnote{\label{foot-dataset}Dataset: }, therefore, the \textit{coverage\_per\_ms} results are visualized as depicted in \fig{vis-list}. For example, ts-config-service has about 83\% coverage while it missed one endpoint our of 6 endpoints. Moreover, this calculation is impacted in \fig{vis-graph} as it shows 3D graph visualization, the color per node which represents the coverage percentage of each microservice in the system.

% Furthermore, the \textit{coverage\_per\_test} is generated and listed in the dataset. It shows that 9 of the tests initiates 25-29 endpoints calls, however, the Login test only initiates 5 endpoints calls, and the Booking test initiates 53 endpoints calls.

% Regarding the \textit{overall\_coverage}, the prototype extracted 262 as all endpoints from the system in the static phase of our methodology and also it produced 119 called endpoints from Selenium tests through the dynamic phase, it produces $119/262\approx 45\%$ as shown in~\tabl{coverage-results}(Methodology Data).

We began the execution by running the deployment script to set up the TrainTicket system on a local instance. Subsequently, our prototype executed the test cases from the provided test benchmark, generated the list of called endpoints and calculated the test coverage according to the described metrics.

The results of the experiment execution revealed a total of 171 unique endpoints extracted from a set of 953 log records generated during the execution of the test cases, out of which 119 endpoints are actual endpoints within the system, 52 endpoints that are related to API-gateway calls. The complete data analysis phases with their results are published in a dataset\footnoteref{foot-dataset}. This dataset contains the complete calculations of $C_{\text{ms}(i)},\ C_{\text{test}(i)}$ metrics.

In terms of evaluating the completeness of our prototype, this case study confirmed that we captured all the endpoints declared in the ground truth. The prototype successfully captured all 262 implemented endpoints in the system, demonstrating the completeness of Stage 1 outcome. For Stage 2 completeness, the prototype extracted all 171 endpoints. Out of the total 171 endpoint calls, our prototype identified 52 distinct calls associated with the API gateway, which are not considered actual endpoints in the system. 

% that were identified in the ground truth during the manual validation of Selenium tests

\begin{table}[b!]
\vspace{-2em}
\renewcommand\arraystretch{1.1}
\centering
% \vspace{-1em}
\scriptsize
%\captionsetup[table]{skip=5pt}
\caption{Summary Statistics of Coverage Metrics} 
\label{tab:coverage-stats-results} 
\begin{tabular}
{@{}p{2cm}|p{2cm}p{2cm}p{2cm}p{2cm}}
\hline
\textbf{Metric} & \textbf{Coverage (\%)} \\
\hline \hline
\textbf{$C_{\text{suite}}$} & 45.42 \\
% \textbf{Activity} & \textbf{\# Endpoints}  \\ \hline
\hline
 & \textbf{Minimum} & \textbf{Average} & \textbf{Maximum} & \textbf{Mode} \\
\hline
\textbf{$C_{\text{ms}(i)}$} & 0 & 44.5 & 100 & 25 \\
\textbf{$C_{\text{test}(i)}$} & 1.14 & 7.29 & 15.27 & 7.25 \\
% Complete system & 262 \\
% Selenium test suite & 119 out of 262\\
% Gatling test suite& 228 out of 262 \\

\hline

% \textbf{Inter-service calls in the test suite   } & \textbf{\# Endpoints}  \\
% \hline
% Manual assessment of inter-service calls in the system & 88 out of 262 \\
% Selenium test suite inter-service call assessment & 68 out of 119 \\
% Gatling test suite inter-service call assessment &  78 out of 228 \\
% Case study inter-service call assessment &  68/171 \\
\hline

\end{tabular}
% \vspace{-2em}
\end{table}

Through the complete data extraction, we calculate the complete test suite coverage to be approximately 45.42\% ($C_{\text{suite}}=\frac{119}{262}\approx 45.42\%$). The summary statistics for the metrics calculations are provided in~\tabl{coverage-stats-results}.

% Through the complete data extraction, we calculate the complete test suite coverage of $\approx 45.42\%$ ($C_{\text{suite}}=\frac{119}{262}\approx 45.42\%$). The summary statistics about the metrics calculations are summarized in~\tabl{coverage-stats-results}.

% we need to highlight examples that achieved the min, max.
% we need to discuss that coverage 100% shows how many endpoint covered

\begin{figure}[t!]
% \vspace{-2em}
\centering
\includegraphics[width=0.55\columnwidth]{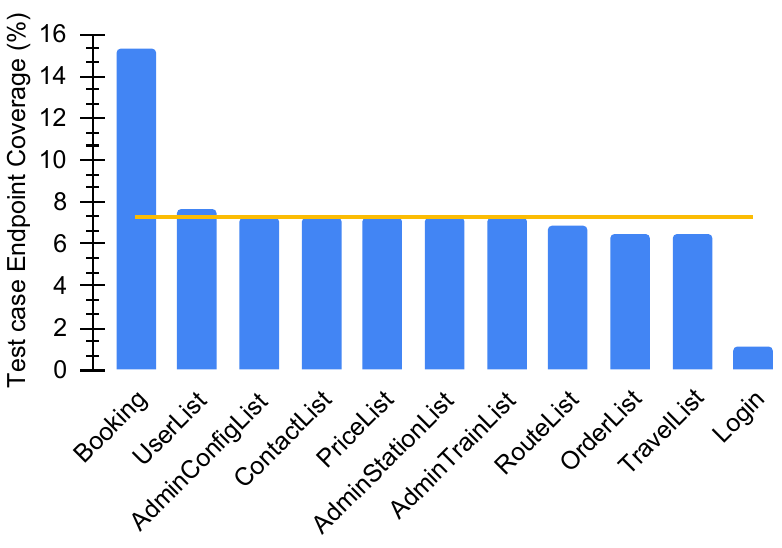}
 \vspace{-1em}
\caption{Test case Endpoint Coverage in the Benchmark Test cases ($C_{\text{test}(i)}$)}
\label{fig:testcase-metric-results}
\vspace{-2em}
\end{figure}

% However, among these 171, our prototype highlighted 52 unique endpoint calls that are related to the API gateway, which are not actual endpoints in the system. 

% However, our approach effectively tests the API gateway components that are not covered by Gatling tests.

% Additionally, our prototype covers an additional 52 endpoint calls related to the API gateway, which are not actual endpoints in the system. However, our approach effectively tests the API gateway components that are not covered by Gatling tests.

 % for both the source code defined endpoints and the Selenium called endpoints. 
% It achieves 100\% coverage of all calls made from Selenium tests. 

The calculation of $C_{\text{test}(i)}$ shows that the maximum coverage achieved by a test case in the study is approximately 15.27\%. This was observed in the \texttt{Booking} test case, which made 53 calls to 40 unique endpoints in the system. On the other hand, the minimum coverage is approximately 1.14\%, which occurred in the \texttt{Login} test case that only called three endpoints. The analysis shows that the average test case endpoint coverage is approximately 7.29\%, while the most common coverage among the test cases is approximately 7.25\%. This coverage was observed in the following five test cases: \texttt{AdminConfigList, ContactList, PriceList, AdminStationList}, and \texttt{AdminTrainList}. \fig{testcase-metric-results} illustrates the endpoint coverage achieved by the 11 test cases, along with the average coverage for better measurement.

% \todo[inline]{what about the other way around microservice by endpoints covered? It could be incomplete = pie chart or table or bar with ..  }

% The calculation $C_{\text{test}(i)}$ shows that $\approx 15.27\%$ is the maximum a test case cover in the study. That happened from the \texttt{Booking} test case that initiated 40 unique endpoint calls. However, the minimum coverage is $\approx 1.14\%$ that occured in \texttt{Login} test case that only calls three endpoints. The analysis shows that the average of test case endpoint coverage is $\approx 7.29\%$, while the most common coverage in the system per test case is $\approx 7.25\%$ from the following five test cases: \texttt{AdminConfigList}, \texttt{ContactList}, \texttt{PriceList}, \texttt{AdminStationList}, and \texttt{AdminTrainList}. The 11 test cases endpoint coverage is depicted in~\fig{testcase-metric-results} along with the average coverage achieved for better measurement.

% most test cases initiate 25-29 endpoint calls, except for the Login test, which only initiates 5 endpoint calls $C_{\text{test}(Login)} = \frac{5}{262} \approx 2\%$, and the Booking test, which initiates 53 endpoint~calls $C_{\text{test}(Booking)} = \frac{53}{262} \approx 20\%$.

\begin{figure}[t!]
% \vspace{-2em}
\centering
\includegraphics[width=.75\columnwidth]{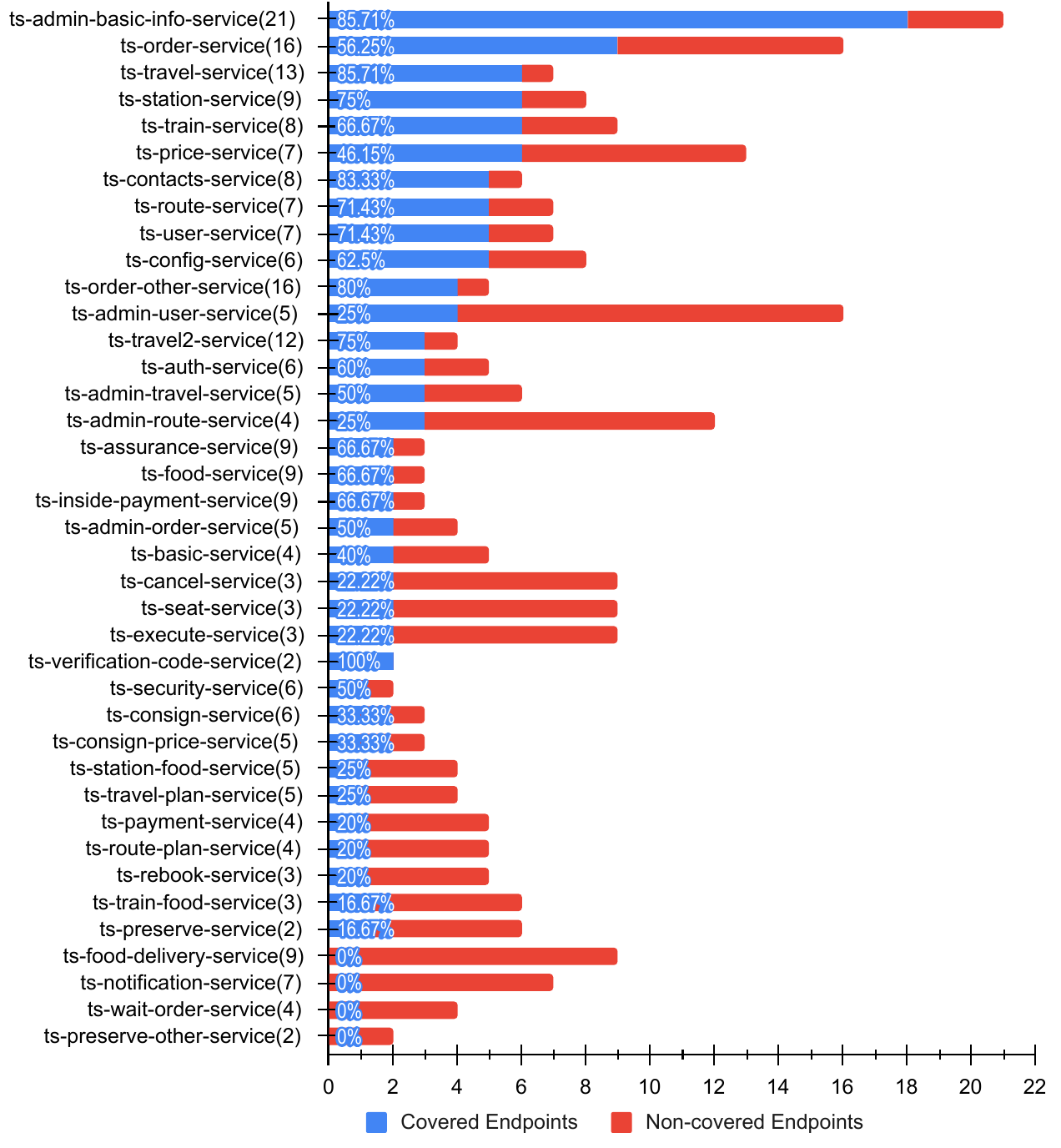}
 \vspace{-1em}
\caption{Microservice Endpoint Coverage in the Benchmark System ($C_{\text{ms}(i)}$)}
{\textit{ The numbers in parentheses indicate the total number of endpoints in each ms.}}
\vspace{-1.5em}
\label{fig:ms-metric-results}
\end{figure}

The calculation of $C_{\text{ms}(i)}$ reveals that the maximum coverage is 100\%, observed in the \texttt{ts-verification-code-service} which has two endpoints covered by the test cases. On the other hand, the minimum coverage is 0\%, indicating that the test suite completely missed testing any endpoints in the following four microservices: \textit{ts-wait-order-service, ts-preserve-other-service, ts-notification-service}, and \textit{ts-food-delivery-service}. The average microservice endpoint coverage is approximately 44.5\%, while the mode statistics show that 25\% is the most common coverage, observed in the following four microservices: \textit{ts-travel2-service, ts-payment-service, ts-route-plan-service}, and \textit{ts-order-other-service}. The complete calculations for each microservice are illustrated in~\fig{ms-metric-results}. 

\begin{figure}[!t]
 \vspace{-0em}
    \centering
       \begin{subfigure}[b]{0.5\columnwidth}
    \centering
       \fbox{\includegraphics[width=0.7\columnwidth]{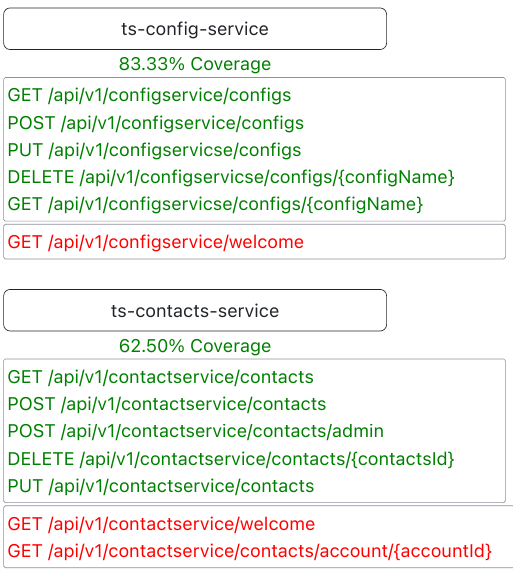}}
    \caption{Microservice endpoint list}
    \label{fig:vis-list}
    \end{subfigure}
    \begin{subfigure}[b]{0.49\columnwidth}
    \centering
      \fbox{\includegraphics[width=0.8\columnwidth]{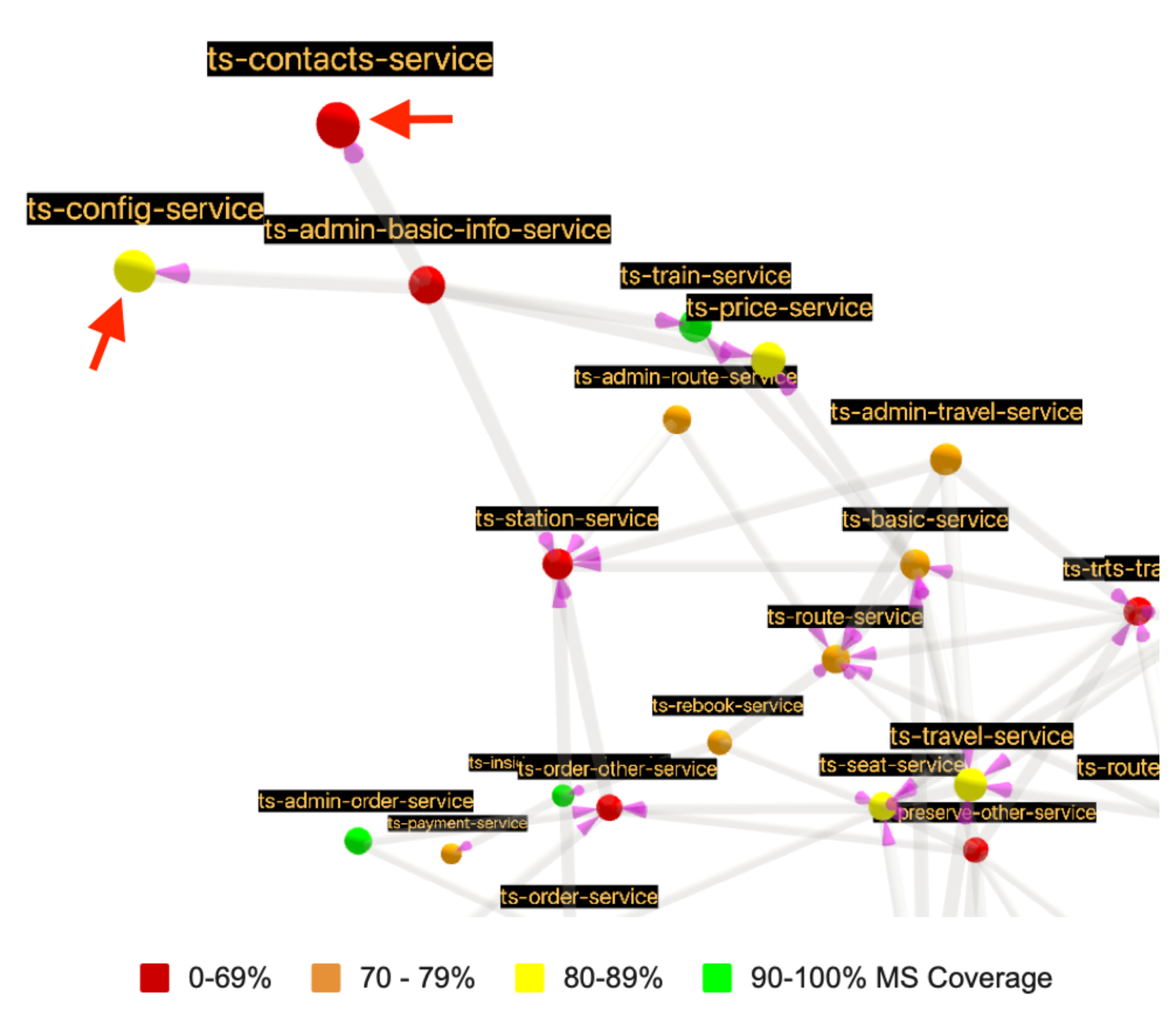}}
    \caption{3D interactive visualizer shows service dependencies (cropped view)} %with coverage color codes ranging from red ($[0-70\%)$), orange ($[70-80\%)$), yellow ($[80-90\%)$) to green ($[90-100\%$])
    \label{fig:vis-graph}
    \end{subfigure}

 \vspace{-2mm}
  \caption{Microservices endpoint coverage visualization (full pictures\footnoteref{foot-prototype-ui})}
   \vspace{-2em}
   % \vspace{-0em}
  \end{figure}

The metrics calculations are visualized using two visualization approaches, as shown in~\fig{vis-list} and \fig{vis-graph}. One with per service view and the other providing the holistic service dependency overview in the context of endpoint coverage. For example, the \texttt{ts-config-service} microservice has an approximate coverage of 83.33\%, missing only one out of six endpoints. This information is also represented in yellow color in the 3D graph visualization, where the color of each node corresponds to the coverage percentage of the respective microservice.

%\todo[inline]{Amr: You need to discuss the Gatling/Static Analysis limitations - The evolution measurements in the proposed method.}
% \vspace{-2em}
\vspace{-1em}
\section{Discussion}
\vspace{-1em}
% \todo[inline]{Amr: We can add discussion about the JMS call extracted from ts-delivery-service}
% Our approach could mitigate E2E test degradation, ensuring continuous system reliability and quality assurance of a decentralized microservice system. 
Our approach has shown promising results in mitigating E2E test degradation and contributing to the continuous reliability and quality assurance of decentralized microservice systems. While further comprehensive data analysis is ongoing, initial findings indicate a positive impact. It determines the log traces connecting tests with endpoints from the current system and a current test suite by automated means. Such traces can help testers manage change propagation as it directly indicates a co-change dependency between specific microservices or endpoints and particular tests. Furthermore, integrating it with CI/CD pipelines would make it an ideal tool to ensure coverage across system evolution changes. On the other hand, it is crucial to consider the context in which the approach is applied, as the user interface may not interact with all middleware endpoints. This can be reflected in the provided metrics, indicating that the E2E test might not achieve 100\% coverage. At the same time, it raises the question of whether the remaining endpoints represent the smell known as \textit{Nobody Home} where the wiring is missing from the user interface, or possibly the endpoints are outdated or dead code. 

% the approach has to be seen in the light where the user interface might not interact with all middleware endpoints, and it might skew provided metrics as the E2E test might not reach 100\% coverage. 

% For instance, when a microservice changes a particular endpoint, E2E testers could be notified of recent changes through issue tracking and trace the tests relevant to the endpoint that needs a review. However, this is left for future work given the limited scope of this work.

It is worth noting that microservices often implement \texttt{isAlive} endpoints for health checks. While some libraries, like Hystrix, can automatically generate these endpoints, some systems implement them manually. As an example, Train-Ticket implemented 39 endpoints that were not utilized in the user interface, rendering them meaningless. Nevertheless, validating these endpoints can guarantee that the system is correctly initialized.

% For example, Train-Ticket implemented 39 endpoints that were not used in the user interface, making them meaningless. However, checking these endpoints before running the test can ensure that the system is properly started.

% It worth noted that microservices might implement \texttt{isAlive} endpoints for health checks, while many libraries (i.e., Hystrix) can automatically generate such endpoints some systems implement them manually. For instance, in Train-Ticket implemented 39 endpoints and did not use them in the user interface, which makes the meaningless. However, using such endpoints checked before the test could assure the system is properly started

 \vspace{-1em}
\subsection{Threats to Validity}
\vspace{-.5em}
% \vspace{-1.5em}
% Case studies often suffer from threats to validity that need to be addressed. We tried to eliminate as many of these as possible that would affect the quality of the study's outcome. 

In this section, we address the potential validity threats to our approach. We adopt Wohlin's taxonomy~\cite{WohlinExperimentation}, which encompasses construction, external, internal, and conclusion threats to validity, as a framework for our analysis.

%\todo[inline]{Tomas: remember if the system has async service like JMS/quarts/timer, our tests traces might show something not triggered by the tests but we assume is run by our tests, we can negate this by running the simulation 2x to find discrepancy or disabling all async calls - difficult}

% \subsubsection{Construction Validity}

% A potential \textbf{construction threat to validity} of our approach is the dependency on static analysis to extract endpoints from the system and dynamic analysis to analyze the centralized traces generated by E2E tests. Therefore, it could pose a threat to our approach if the system under test lacks the required source code for analysis or if the source code does not adhere to common standards for endpoint definition. Additionally, if the system lacks support for centralized traces that can be captured and analyzed, it could also impede~our~approach.

A potential \textbf{construction validity threat} arises from the dependency on static analysis for endpoint extraction and dynamic analysis of centralized traces generated by E2E tests. It includes missing or non-standard source code and a lack of support for centralized traces, which can hinder our approach.

% This threat could materialize if the system under test lacks the necessary source code or if the source code does not follow standard endpoint definition practices. Furthermore, if the system does not support centralized traces that can be captured and analyzed, it may hinder our approach.

% does not provide the necessary source code to be analyzed or if the source code does not follow specific standards for defining endpoints. And also, if the system does not support centralized traces that can be captured~and~analyzed.

Our prototype is currently implemented for specific programming languages and frameworks. However, it is important to note that the methodology itself is not limited to these specifications. It can be adapted and applied to other languages and frameworks, mitigating construction threats related to dependencies. Moreover, asynchronous messaging poses a potential risk to test execution by causing ghost endpoint call trace events. To mitigate this threat, potential approaches include disabling asynchronous services or conducting repeated test executions to minimize the impact.

\textbf{Internal validity threats} arise from potential mismatches between the extracted endpoint signatures from the source code and the traces. Although overloads are infrequent, inaccurate matching may occur due to trace values not aligning precisely with the defined types in the code. For example, if a trace contains an integer in the URL, it may match with an integer parameter type even if the corresponding endpoint has a string parameter type. Moreover, Multiple authors collaborated to ensure accurate data and calculations. They independently verified and cross-validated the results, rotating across validation processes to minimize learning effects.

To address \textbf{external validity threats}, our case study utilized a widely recognized open-source benchmark to evaluate its endpoints coverage using our proposed approach. Still, it is important to acknowledge that the results and conclusions drawn from this specific benchmark may not fully represent the entire range of microservices systems that adhere to different standards~and~practices.

One potential \textbf{conclusion validity threat} is that our tool was tested on an open-source project rather than an industry project. However, we aimed to address this by selecting an open-source project that employed widely-used frameworks in the industry. Furthermore, to ensure the reliability and consistency of our results, we performed the case study in multiple environments and confirmed that the outcomes remained consistent.

% One \textbf{conclusion validity threat} is that our tool was tested only on an open-source project rather than an industry project. However, we attempted to mitigate this threat by selecting an open-source project that utilized frameworks commonly used in the industry. Additionally, to enhance the reliability and consistency of our results, we re-executed the case study in different environments and verified that the outcomes remained consistent.

%\todo[inline]{Amr: Please add the threats of concluding the results of the study.}

\section{Conclusion}
\vspace{-1em}

Despite the broad adoption of microservices for software solutions, there are open challenges practitioners face with E2E testing. While testers might assume complete test coverage, verification mechanisms on the actual state of test completeness within the system are missing.
We sought to define metrics and establish an approach to calculate the E2E test suites coverage of microservice system endpoints.  
Our approach determines the connection between individual tests and microservice endpoints, which are the system entry points for user interfaces used by E2E testers. We performed a case study on an established system benchmark and a test suite aiming for full coverage, revealing that the achieved coverage fell significantly short of being comprehensive.
% We performed a case study on an established system benchmark and a test suite that aimed for full coverage demonstrating the coverage is far from comprehensive. 

In future work, we will explore system and test suite evolution, evaluating how our approach guides co-coupling between system changes and tests to ensure quality assurance and reduce test suite degradation. We also plan to expand our metrics to encompass different test paths within the endpoints.

% Future work will experiment with system and test suite evolution, assessing our approach guidance on co-coupling between system changes and tests to maintain high-quality assurance and mitigate test suite degradation. We also aim to broaden our metrics to recognize various test paths within the endpoint.

\vspace{-0.5em}
\section*{Acknowledgements}
\vspace{-6px}
This material is supported by the National Science Foundation under Grant No. 2245287 and Grant No. 349488 (MuFAno) from the Academy of Finland.

% This material is based upon work supported by the National Science Foundation under Grant No. 2245287, and a Grant No. 349488 (MuFAno) from the Academy of Finland.

\vspace{-0.5em}

\bibliographystyle{splncs04}
\bibliography{mybibliography}




% LNCS Homepage, \url{http://www.springer.com/lncs}. Last accessed 4
% \end{thebibliography}

    % \theendnotes

\end{document}